\documentclass{article} 
\usepackage{epsfig}
\title{\bf Quantum-Critical Dynamics of the Skyrmion Lattice.} 
\author{A. G. Green} 
\begin{document}
\maketitle 
\begin{abstract}
Near to filling fraction $\nu=1$, the quantum Hall 
ferromagnet contains multiple Skyrmion spin excitations. This multi-Skyrmion 
system has a tremendously rich quantum-critical 
structure. This is simplified
when Skyrmions are pinned by disorder. We calculate 
the nuclear relaxation rate in this case and compare the result with experiment. 
We discus how such measurements may be used to further probe the quantum-critical 
structure of the multi-Skyrmion system.
\end{abstract}
\vspace{0.25in}

At exact filling of a single Landau level the quantized Hall
state forms an almost perfect ferromagnet. This quantum Hall ferromagnet (QHF)
has some novel features due to the phenomenology of the underlying quantized 
Hall state. 
Magnetic vortices, or Skyrmions, in the QHF carry quantized 
electrical charge\cite{Sondhi}. These Skyrmions are stabilised by a chemical 
potential so that the ground state slightly away from filling fraction 
$\nu=1$  contains a finite density of them. 
This original proposal of Sondhi {\it et al.} has been 
confirmed in a number of experiments\cite{Sondhi,Experiment}. 

The $T=0$ phase diagram of the multi-Skyrmion system has been 
thoroughly investigated. In the absence of disorder, crystalline 
arrangements are expected\cite{Crystals1,Abolfath1}. 
For filling
fractions very close to $\nu=1$ a triangular lattice is formed with 
a transition to a square lattice as the deviation from
$\nu=1$ is increased. The statistical mechanics of the possible 
melting transitions has been considered\cite{Timm_melt}. 
At the highest Skyrmion densities, zero point 
fluctuations are expected to give rise to a quantum-melted
state\cite{Quantum_melt}.

Despite this wealth of study, a complete account of the experimental 
observables has not been achieved. For example, nuclear
magnetic resonance provides one of the clearest probes of the spin polarisation
in the QHF\cite{Experiment}. Although it is understood in general terms how 
low-energy 
spin-fluctuations of the Skyrmion system may enhance the  relaxation of 
nuclear spins, 
attempts to calculate relaxation rates have been flawed\cite{Cote}.
The fundamental physics, missed in other considerations,
is the quantum-critical nature
of fluctuations of the Skyrmion lattice. One immediate consequence of this
quantum-criticallity is that the limits of temperature/frequency$\rightarrow0$
and frequency/temperature$\rightarrow0$ are very different. Typically, 
experimental probes are at frequencies much less than temperature and the
latter limit is appropriate. This means that zero-temperature calculations 
{\it cannot} model experiments correctly\cite{Similar}.

Let us consider these points a little further. In our analysis below, we will
find an underlying gappless XY-model governing orientational fluctuations
of the multi-Skyrmion system. If a spinwave expansion is 
attempted for a gappless magnet at or below its critical dimension, the 
occupation of low-frequency modes is found to diverge. The constraint, 
fixing the 
magnitude of the local spin, restricts this divergence. Interplay between
divergence and constraint gives rise to a finite (temperature dependent) 
correlation length, $\xi(T)$, beyond which correlations of the magnet decay 
exponentially. The dynamics of the critical magnet are very different on 
length scales greater or less than $\xi$.
On lengthscales less than $\xi$, the groundstate is ordered (albeit
in a quantum superposition of all possible orientations due to long
wavelength spin fluctuations). Fluctuations with wavelength less than $\xi(T)$ 
may, therefore, be described by a modified spinwave expansion. 
On lengthscales greater than $\xi$, the 
groundstate is disordered and fluctuations are overdamped. This 
{\it quantum relaxational dynamics} is a striking feature of quantum-critical 
systems and leads to interesting universalities\cite{Read}. 
Regimes of renormalized classical and quantum activated behaviour
at low-temperature, cross over to universal behaviour in the 
high-temperature quantum-critical regime.

 The Skyrmion spin-configuration consists of a vortex-like 
arrangement of in-plane components of spin with the z-component reversed 
in the centre of the Skyrmion and gradually increasing to match the 
ferromagnetic background at infinity. At large distances, the spin 
distribution 
decays exponentially to the ferromagnetic background on a length scale 
determined by the ratio of spin stiffness to Zeeman energy\cite{pure_Skyrmion}.
An individual 
Skyrmion may be characterized completely by its position ({\it i.e.} the 
point at which the spin points in the opposite direction to the ferromagnetic
background) its size ({\it i.e.} the number of flipped spins) and the 
orientation of the in-plane components of spin. The equilibrium size of the Skyrmion 
is determined by a balance between its coulomb and Zeeman energies\cite{Sondhi}
(In the presence of a disorder potential, the potential
energy of the Skyrmion also enters this balancing act\cite{Disorder}).

Consider a ferromagnet with a dilute distribution of Skyrmions.
The normal modes of this system are relatively
easy to identify. Firstly, ferromagnetic spinwaves propagate in-between
the Skyrmions. The spectrum of these is gapped by the Zeeman
energy and will be ignored from now on.
Positional fluctuations, or phonon modes, of the Skyrmions are
gapless in a pure system, but gapped when
the lattice is pinned by disorder.  Finally, fluctuations in the in-plane orientation and 
size must be considered. 
These two types of fluctuation are intimately connected; rotating a
Skyrmion changes its size. This follows from the
commutation relations of quantum angular momentum 
operators\cite{Abolfath1,Cote,Nazarov,Piette}.

The orientation, $\theta({\bf x}_i,t)$, of a Skyrmion centred 
at a point ${\bf x}_i$ is described by the 
following effective action\cite{Cote,Nazarov}:
\begin{equation}
S=\frac{1}{2}
\int dt 
\left[
\sum_i 
I_i
\theta({\bf x}_i,t) \partial_t^2\theta({\bf x}_i,t)
- \sum_{<i,j>} 
J_{ij}\cos (\theta({\bf x}_i,t)-\theta({\bf x}_j,t))\right].
\label{Theta_action}
\end{equation}
$I_i$ is the moment of inertia of the $i^{th}$ Skyrmion and $J_{ij}$ is the 
stiffness to relative rotations of neighbouring Skyrmions. The first term
in Eq.(\ref{Theta_action})
arises due to the change in energy of a Skyrmion when its size fluctuates;
$\Delta E=I^{-1} \delta s^2/8$. Since the z-component of spin and orientation 
are conjugate coordinates, a cross-term 
$i\delta s({\bf x}_i,t) \partial_t \theta({\bf x}_i,t)/2$
appears in their joint effective action. Integrating out 
$\delta s({\bf x}_i,t) $ gives
the first term in Eq.(\ref{Theta_action}). 
Clearly, $1/4I$ is the second derivative of the Skyrmion energy with 
respect to its spin. $I$ is related to the Skyrmion size 
(in fact $I=24\mu_BgB/s$\cite{Abolfath1}) and, in the absence of disorder, is 
the same for all Skyrmions.
Correlation functions involving 
$\delta s$ may be calculated using Eq.(\ref{Theta_action}) by making the 
replacement 
$\delta s({\bf x}_i,t) \rightarrow 2i I \partial_t \theta({\bf x},t)$-
the result of a simple Gaussian integration over $\delta s$.
 The second term in Eq.(\ref{Theta_action}) is an effective dipole-interaction of Skyrmions
due to the energetics of overlapping Skyrmion tails\cite{Abolfath1,Piette}.
In a square Skyrmion lattice\cite{hexagonal}, $J_{ij}$ is independent of 
lattice site.  
A continuum limit may be taken where $\theta_i$ is replaced by a staggered
field, $\theta_i \rightarrow \theta_i +\eta_i \pi$, with $\eta_i=0,1$ on 
adjacent sites, and $\theta_i-\theta_j$ is replaced by a derivative;
\begin{equation}
S= \frac{1}{2}
\int \frac{d\omega d^2k }{(2 \pi)^3}
\theta({\bf k},\omega)
\left(
\delta \nu \bar  \rho I \omega^2
+
J{\bf k}^2
\right)
\theta(-{\bf k},-\omega).
\label{Theta_action2}
\end{equation}
The frequency integral in this expression is shorthand for a 
Matsubara summation at finite temperature and the momentum integral is over 
the Brillouin zone\cite{Brillouin_zone}.
A factor of the Skyrmion density $\delta \nu \bar \rho$ has been introduced, 
where 
$\delta \nu$ is the deviation from filling fraction $\nu=1$ and
$\bar \rho$ is the electron density.
There are a few caveats to the use of Eq.(\ref{Theta_action2}). We defer
discusion of these until later. 
This model is perhaps most familiar as an effective theory of the Josephson 
junction array. In this case, $\theta$ is the phase of the superconducting
order parameter and its conjugate coordinate is the charge of the 
superconducting junction. 

In order to calculate properties of the Skyrmion lattice, we must relate 
fluctuations in the Skyrmion orientation to fluctuations in the orientation
of local spin. We use a coherent-state representation of the polarization of 
the local spin, {\it via} an O(3)-vector field ${\bf n}({\bf x},t )$. The 
static spin distribution at a point ${\bf x}$ relative to the centre of a 
single Skyrmion is denoted by ${\bf n}({\bf x})$
and its in-plane components by $n_x+in_y=n_r  e^{i \phi_0}$. The in-plane 
components of local spin at a point ${\bf x}$, in response to rotational
fluctuations of a Skyrmion centred at ${\bf x}_i$, are given by 
\begin{equation}
n_r e^{i\theta}({\bf x},t)=
(-)^{\eta_i}
\left[ n_r({\bf x}-{\bf x}_i)
+
2iI \frac{\partial n_r({\bf x}-{\bf x}_i)}{\partial s}  \partial_t \theta({\bf x}_i,t)
\right]
e^{i\theta+i\phi_0}({\bf x}_i,t).
\label{in_plane_component}
\end{equation}
We have used the conjugate relationship between Skyrmion spin and orientation 
in writing down this expression.
In a distribution of many Skyrmions, one must in principle sum the 
contributions of all Skyrmions to the fluctuation in spin at the point ${\bf x}$. However, in the dilute limit in which we are performing our explicit 
calculation, Skyrmions are exponentially localized. The dominant spin 
fluctuations occur near to the centre of Skyrmions and so, to logarithmic
accuracy, the local fluctuations at a point ${\bf x}$ are due only to 
the nearest Skyrmion.

We will calculate the
nuclear relaxation rate due to low energy quantum fluctuations of the Skyrmion
lattice;
\begin{equation}
\frac{1}{T_1}
=
T \gamma  \lim_{\omega \rightarrow 0}
\int \frac{d^2k}{(2 \pi)^2}
\frac{
{\cal I}m
\langle 
S_+({\bf k}, \omega) S_-(-{\bf k}, -\omega)
\rangle
}{\omega},
\label{nuclear_relaxation}
\end{equation}
where $\gamma$ is the hyperfine coupling constant. Other physical observables,
such as the temperature dependence of magnetization,
$
\langle M \rangle
= 
\int dt
d^2x \langle n_z({\bf x},t) \rangle,
$
may be calculated similarly.
Our first task is to replace the expectation of the spin raising and lowering
operators in Eq.(\ref{nuclear_relaxation}) by correlators of the Skyrmion
orientation. Substituting from Eq.(\ref{in_plane_component}) into
Eq.(\ref{nuclear_relaxation}) and ignoring terms higher order in frequency,
the nuclear relaxation rate at a point ${\bf x}$ is
\begin{equation}
\frac{1}{T_1}({\bf x})
=
T \gamma  \lim_{\omega \rightarrow 0} n_r^2({\bf x})
\int \frac{d^2k}{(2 \pi)^2}
\frac{
{\cal I}m
\langle 
e^{i\theta}({\bf k}, \omega) e^{-i\theta}(-{\bf k}, -\omega)
\rangle
}{\omega}.
\label{nuclear_result}
\end{equation}
This takes the form of a correlation function of the Skyrmion orientation, 
multiplied by a profile function characteristic of the Skyrmion groundstate.
The average rate is given by integrating this  over an area containing a single
Skyrmion and multiplying by the Skyrmion density $\delta \nu  \bar \rho$. The 
result is identical to Eq.(\ref{nuclear_result}) with the replacement 
$n_r^2({\bf x}) \rightarrow  \delta \nu \overline{n^2_r}$.
$\overline{n_r^2}= \bar \rho \int d^2x n_r^2({\bf x})$ is a number characteristic
of a single Skyrmion. For a pure Skyrmion spin distribution\cite{pure_Skyrmion},
 $\overline{n_r^2}=2s$ 
to logarithmic accuracy in the Skyrmion spin, $s=\bar \rho \int d^2x (1-n_z({\bf x}))$. 
Notice that radial fluctuations of local spin contribute only to higher order
in frequency and have been neglected in writing down Eq.(\ref{nuclear_result}). 

The problem of finding the nuclear relaxation rate has now been reduced to 
evaluating the correlation function in Eq.(\ref{nuclear_result}) using the
effective action Eq.(\ref{Theta_action2}). This is rather tricky. 
The O(2)-quantum
rotor, described by Eq.(\ref{Theta_action2}), is quantum-critical. 
It is necessary to employ a non-perturbative scheme, such as
1/N or epsilon expansions, to calculate in the quantum-critical regime of 
this model.
Here we use the result of the 1/N expansion of 
Chubukov {\it et al}\cite{Chubukov}.

An important feature of the effective action, Eq.(\ref{Theta_action2}), is 
that it displays
a zero-temperature phase transition. 
For $IJ<1$, the Skyrmion moment of inertia is sufficiently small that 
quantum fluctuations destroy long range order even at zero temperatures. 
For $IJ>1$, the $T=0$
groundstate has an infinite correlation length  and is 
ordered. Notice
that arbitrarily small temperatures destroy this long-range order even 
when $IJ>1$. This arises due to the interplay between fluctuations 
and constraint and can
be seen in a simple mean-field calculation\cite{Chubukov}. 
In the O(2) representation of Eq.(\ref{Theta_action2}),
$S=\int dt d^2x \left[ {\bf n} (I\partial_t^2+ J\partial_{\bf x}^2 ){\bf n} 
+\lambda({\bf x},t) ({\bf n}^2-1) \right]$, where ${\bf n}({\bf x},t)$ is 
an O(2)-vector
field and $\lambda({\bf x},t)$ is an auxiliary field that imposes the 
constraint
${\bf n}^2=1$. Imposing the constraint at mean-field level, 
$\langle {\bf n}^2 \rangle=1$,
determines a temperature dependent gap, $\lambda(T)$.
The spin correlations decay exponentially on a length
scale $\xi(T)=\sqrt{J/\lambda(T)}$. 
The results of such a calculation are sketched in Fig.1.
\begin{figure}
\centering
\epsfig{file=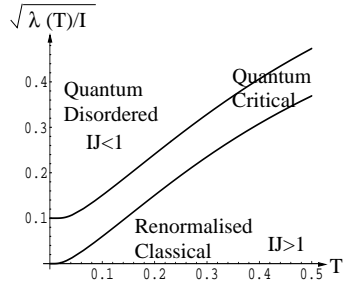,height=1.5in}
\caption{T-dependence of the Gap.}
\end{figure}
\begin{figure}
\centering
\epsfig{file=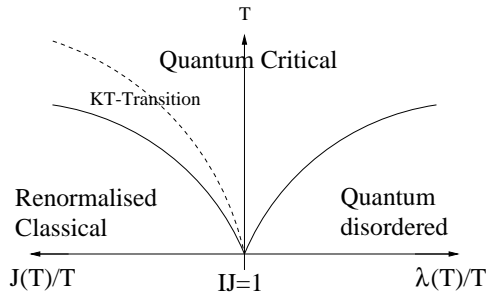,height=1.5in}
\caption{Phase diagram of the O(2)-vector model.}
\end{figure}
 Above a temperature of about 
$T_{QC}=|2 \pi J-E_{\hbox{max}}|$ 
(The cut-off, $E_{\hbox{max}}=2 \pi \sqrt{J/I}$ corresponds to fluctuations 
with momentum at the Brillouin zone boundary, 
$k_{\hbox{max}}=\pi \sqrt{\bar \rho \delta \nu}$\cite{Brillouin_zone}), the
gap/correlation length develops a universal temperature dependence. In this 
region, thermal and quantum fluctuations are of similar importance and are very difficult
to disentangle. This crossover to universal high-temperature behaviour from distinct 
low-temperature behaviours is a feature of all correlation functions of 
Eq.(\ref{Theta_action2}). It is usual to summarise this behaviour by the phase diagram sketched in
Fig.2. In this figure, $J(T)$ is the renormalized 
stiffness in the ordered phase and $\lambda(T)$ is the gap in the paramagnetic phase.

The correlation function required in Eq.(\ref{nuclear_result}) has been 
calculated in Ref.\cite{Chubukov} by means of a 1/N expansion, with the 
result
\begin{eqnarray*}
& &
\;\;\;\;
\lim_{\omega \rightarrow 0}\frac{1}{\omega}
\int \frac{d^2k}{(2 \pi)^2} 
{\cal I}m
\langle
e^{i \theta}({\bf k}, \omega) e^{-i\theta}(-{\bf k}, -\omega) \rangle
\\
& &
\begin{array}{llll}
= &
 \frac{1}{T} \frac{0.015}{\sqrt{IJ}} 
\left( \frac{kT}{\pi J} \right)^{\eta},
&
\mbox{\scriptsize $T \gg T_{QC}$} &
\\
= &
\frac{1}{T}
\frac{0.085}{\sqrt{IJ}}
 \frac{IT^2}{\lambda(0)} e^{-2 \sqrt{\lambda(0)/IT^2}},
&
\mbox{\scriptsize $T \ll T_{QC}$,} &
\mbox{\scriptsize  $IJ<4$},
\\
= &
\frac{0.18}{T}
 \left( \frac{kT}{2\pi \lambda(T)} \right)^{1/2},
&
\mbox{\scriptsize $T \ll T_{QC}$,} &
\mbox{\scriptsize  $IJ>4$},
\end{array}
\end{eqnarray*}
where $\eta$ is a number close to zero. 
Substituting these results into Eq.(\ref{nuclear_result}) we obtain
\begin{equation}
\frac{1}{T_1}=
\gamma \frac{0.03s}{\sqrt{IJ}} 
\left( \frac{kT}{\pi J} \right)^{\eta}
\end{equation}
at high temperature. The full behaviour is sketched in Fig.3. The kink at $T_{KT}$
is due to the discontinuous change in spin stiffness seen at the Kosterlitz-Thouless transition. This effect is not seen in the 1/N expansion of Ref.\cite{Chubukov} and must be calculated by some other means.
\begin{figure}
\centering
\epsfig{file=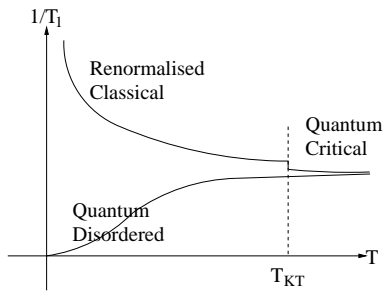,height=1.5in}
\caption{Temperature of the nuclear relaxation rate, $1/T_1$.}
\end{figure}

The nuclear relaxation rate obtained here is very different from that 
obtained in Ref.\cite{Cote}. As pointed out in the introduction, this is 
due to the unphysical limit $T/\omega \rightarrow0$ used in Ref.\cite{Cote}.
Nevertheless, it is instructive to see how the results of Ref.\cite{Cote}
relate to the present formalism. Since $T\ll \omega$ in their work, 
C\^ot\'e {\it et al}
consider fluctuations on length scales much less than the correlation length.
The groundstate is ordered and a spinwave expansion may be used. Long
wavelength fluctuations lead to a superposition of orientations- rotational
averaging- the immediate consequence of which is that 
$\langle e^{i\theta}({\bf k},\omega)e^{-i\theta}(-{\bf k},-\omega)\rangle=0$.
Returning to the substitution of Eq.(\ref{in_plane_component}) into 
Eq.(\ref{nuclear_relaxation}), we must retain terms to next order in 
frequency. This is a cross-term between radial and transverse 
fluctuations and involves a correlator 
$\langle \partial_t \theta
e^{i\theta}({\bf k}, \omega) e^{-i\theta}(-{\bf k}, -\omega)
\rangle$.
Evaluating this correlation function {\it via} a $T=0$ 
spinwave expansion of the effective action, Eq.(\ref{Theta_action2}), 
reproduces the
result of Ref.\cite{Cote} (up to a numerical factor due to our estimate of the 
Skyrmion profile function, $n_r \partial n_r/\partial s$). 
The zero-temperature phonon contribution may
be calculated similarly.

Up to now, we have assumed that the Skyrmion lattice is pinned by disorder
 and that phonons may be ignored as a consequence. The situation is rather
subtle and a fuller discussion is appropriate at this juncture.
A quadratic effective action for phonons of the Skyrmion lattice is 
known\cite{Crystals1,Abolfath1,Timm_melt}. It is identical to that of an
electronic Wigner crystal in a magnetic field with a vanishing effective 
mass\cite{mass}. Under this effective action, Skyrmions move in small ellipses
with a frequency $\omega_{\bf k}\sim |{\bf k}|^{3/2}/B$ and major axes 
orientated transverse to the phonon momentum.

At finite temperature, the occupation of transverse phonons is infra-red 
divergent. This divergence is restricted 
by
interactions between phonons arising from the non-harmonicity of the 
Skyrmion interaction. Unlike fluctuations in orientation, where the 
spinwave interaction is due to a topological constraint, these phonon 
interactions are non-universal (at best the universality is hidden in the  
details of the groundstate spin distribution and effective Skyrmion interaction
potential). The resulting physics is very similar to that discussed for the 
rotation mode above; the phonon system is quantum-critical and has 
low-temperature ordered and quantum-melted\cite{Quantum_melt} phases and a 
high-temperature quantum-critical regime.

Even this is not the full story. Although the low energy dispersions of
phonons and Skyrmion rotations are independent\cite{Cote}, non-linear interactions exist 
between these modes. The lattice stiffness is, in part, due to the dipole interaction of Skyrmions 
and is affected by fluctuations in orientation.
Similarly, the dipole interaction between Skyrmions is strongly dependent upon the 
separation of Skyrmions and is affected by phonons. These non-linearities
occur on the same footing as the phonon-phonon interactions and 
rotation-rotation interactions. The full quantum-critical structure of the 
multi-Skyrmion system is tremendously complicated. 

The position taken here in neglecting this wealth of structure is that
the Skyrmion lattice is pinned by disorder and the phonon spectrum gapped. 
Phononic fluctuations are suppressed at low temperatures and the associated critical structure occurs 
at higher temperature. The residual effect of phonons is 
a slight thermal renormalization of the rotational stiffness.
The slight distortion in static positions of Skyrmions, in response to the
disorder potential, gives a small random contribution to the stiffness $J$.
This randomness produces a small region of Bose-glass
phase at low temperatures, intervening between the paramagnet and renormalized
classical regimes. For weak disorder, this phase only affects the physics very
close to the critical point and does not affect our conclusions.

We now turn to a discussion of the experimental implication of the above 
calculations. Detailed measurements of $1/T_1$  have been carried out by Bayot 
{\it et al}\cite{Bayot}. Above $40mK$, $T_1$ is independent of 
temperature. This is consistent with the rotational degrees of freedom 
being in their quantum-critical regime. Values of $I$ and 
$J$ for this system extracted from TDHFA calculations\cite{Cote} 
put $IJ$ very close to $1$. The system is close to criticality and 
the crossover to the quantum-critical regime occurs at correspondingly 
low temperature.
At $40mK$ there is an abrupt step in $1/T_1$(and attendant peak in 
heat capacity). This is consistent with a 
Kosterlitz-Thouless transition in the orientational order (notice that
the crossover temperature $|2 \pi J-E_{\hbox{max}}|$ may be much less
than $T_{KT}=2 \pi J$ and so the behaviour may be quantum-critical
either side of the transisiton). There are a
number of other candidate transitions\cite{Crystals1,Timm_melt,Cote},
however, and it is not easy to discriminate between them. Considerations
along the lines of those presented here allow some elaboration, but
this is necessarily rather speculative and we refrain from its discussion
at present.
We may make some firm predictions for  nuclear
relaxation measurements below $40mK$. Changing the deviation in filling
fraction or using tilted filed measurements both change the parameter
$IJ$ and allow exploration of the phase diagram shown in Fig.1. The divergence
or otherwise of the nuclear relaxation rate as temperature is reduced to zero
should give a clear indication of the quantum-critical structure.

I would like to thank N. R. Cooper,  J. R. Chalker, S. M. Girvin and 
N. Read for enlightening discusions, comments and suggestions. 
This work was supported by Trinity College Cambridge.


\begin{thebibliography}{99}
\bibitem{hexagonal}In the case of the hexagonal lattice,
anti-ferromagnetic order is frustrated and a different description is required.
\bibitem{Brillouin_zone} The Skyrmion lattice has an anti-ferromagnetic 
ordering of in-plane spins. The unit cell, therefore, contains two Skyrmions
and the Brillouin zone must be chosen appropriately.
\bibitem{mass} A curious feature of the nearly filled Landau level is that
quasi-particles have only a magnus-force term in their effective action and no
kinetic term.
\end{thebibliography}
\end{document}